# Nonlinearity tunes crack dynamics in soft materials

Fucheng Tian[a*], Jian Ping Gong [a, b†]

[a] *Faculty of Advanced Life Science, Hokkaido University, Sapporo 001-0021, Japan;*
[b] *Institute for Chemical Reaction Design and Discovery (WPI-ICReDD), Hokkaido University, Sapporo 001-0021, Japan*

Cracks in soft materials exhibit diverse dynamic patterns, involving straight, oscillation, branching, and supershear fracture. Here, we successfully reproduce these crack morphologies in a two-dimensional pre-strained fracture scenario and establish crack stability phase diagrams for three distinct nonlinear materials using a fracture phase field model. The contrasting phase diagrams highlight the crucial role of nonlinearity in regulating crack dynamics. In strain-softening materials, crack branching prevails, limiting the cracks to sub-Rayleigh states. Yet strain-stiffening stabilizes crack propagation, allowing for the presence of supershear fracture. Of particular interest is the large-strain linear elastic materials, where crack oscillation is readily triggered. The onset speed of such instability scales linearly with the characteristic wave speed near the crack tip, supporting the notion that such crack oscillations are a universal instability closely tied to the wave speed. The oscillation wavelength is shown to be a bilinear function of the nonlinear scale and crack driving force, with a minimum length scale associated with the dissipative zone. Moreover, our findings suggest that the increase in characteristic wave speed due to strain-stiffening can account for the observed transition of cracks from sub-Rayleigh to supershear regimes.

[*] Correspondence author: tianfc@sci.hokudai.ac.jp (Fucheng Tian)
[†] Correspondence author: gong@sci.hokudai.ac.jp (Jian Ping Gong)

Fracture, as a prime avenue for material failure, has ignited enduring interest in the realms of materials science, physics, and nonlinear mechanics [1-4]. Due to the presence of large-strain nonlinearities, the fracture of soft materials exhibits heightened complexity, giving rise to diverse dynamic morphologies and intriguing symmetry-breaking instabilities [5-8]. In brittle gels with slight strain softening, cracks subjected to tensile loading (Mode-I) undergo microbranching instability in three dimensions at a critical speed of $v_c > 0.4c_R$ ( Rayleigh wave speed, $c_R$ ) [9-11]. As the system approaches quasi two-dimensional (2D) states, the crack exhibits path oscillations and tip-splitting at a higher critical speed of $v_c > 0.9c_s$ (shear wave speed, $c_s$ ) [12-14], forbidding its speed beyond $c_R$ ($c_R \approx 0.933c_s$). Whereas in rubbery materials characterized by strain stiffening, the crack spreads more stably [15,16], with few bifurcations or oscillations. Moreover, the crack speed in such materials can readily exceed its $c_s$, resulting in supershear fracture [17-19]. These intricate crack dynamics transcend the scope of classical Linear Elastic Fracture Mechanics (LEFM) and are thought to be closely related to the nonlinearity and characteristic scales near the crack tip [1,20-22].

Recent experiments and simulations have highlighted two characteristic length scales that are absent in LEFM [23-26], i.e., nonlinear scale $\ell_n$ and dissipative scale $\xi$, see Fig. 1(a). $\ell_n$ arises from the nonlinearity at the crack tip where LEFM breaks down [14], scaling with the ratio of fracture energy ($\Gamma_0$) to shear modulus ($\mu_0$). $\xi$ represents the intrinsic dissipative scale near the crack tip, where the elastic energy transported from the remote parts by wave is dissipated to form new crack surfaces [6,26]. The most encouraging outcome to date in supplementing these two scales is the unveiling of the scaling laws governing finite wavelength in crack oscillation instabilities [24,27]. As per the prevailing view, such crack oscillation is a universal phenomenon, known as supercritical Hopf bifurcation [25,28]. To the best of our knowledge, however, this instability pattern has not yet been captured in a broader range of soft materials beyond brittle gels, such as rubber-like elastomers. The disparity in crack dynamics reminds us of their distinct nonlinear responses, thereby posing an open issue: how does the nonlinearity of soft materials tune their crack dynamics?

In this context, this work aims to address the aforementioned open issue. Using the latest developed dynamic phase-field model (PFM) [24,29], we constructed crack stability phase diagrams for three diverse nonlinear materials in a 2D pre-strained fracture scenario, clarifying the rare occurrence of crack oscillation instability in experiments. The instability wavelength is identified as a bilinear function of nonlinear scale and crack driving force, featuring an intrinsic minimum scale. The onset speed of oscillation scales linearly with the characteristic wave speed near the crack tip. Moreover, our findings also suggest the transition of cracks from sub-Rayleigh to supershear regimes in homogeneous soft materials roots in the increased characteristic wave speed caused by strain stiffening.

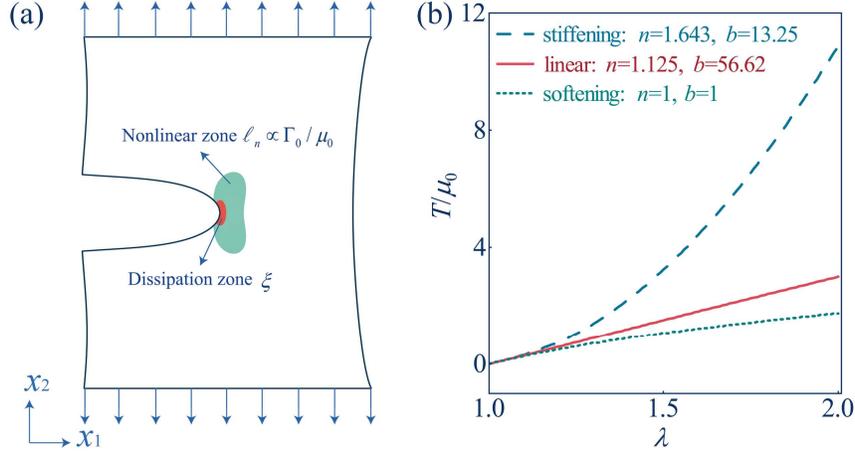

**FIG.1** (a) Illustration of two intrinsic characteristic scales near the crack tip under tensile loading (Mode-I). (b) Curves of normalized nominal stress $T/\mu_0$ ($\mu_0$ is the initial shear modulus) versus uniaxial stretch ratio $\lambda$ depicted by the generalized neo-Hookean (GNH) model: strain stiffening ($n=1.643$, $b=13.25$), large strain linear elasticity ($n=1.125$, $b=56.62$), and strain softening ($n=1$, $b=1$).

The present topic involves crack dynamics, for which we establish dynamic fracture scenarios using pre-strained crack configurations that closely resemble the experimental setup [2,30]. Firstly, a pristine rectangular sheet of width $w_0$ and height $h_0$ is stretched to a specified strained state. Then maintaining this strained state, a notch is created by a sharp blade to trigger dynamic fracture (more details can be found in the Supplementary Materials [31]). The mechanical response of this sheet is postulated to be described by the plane-stress generalized neo-Hookean (GNH) model with the following strain-energy function [32-34]

$$W(\lambda) = \frac{\mu_0}{2b}\left\{\left[1+\frac{b}{n}\left(\lambda_1^2 + \lambda_2^2 + \lambda_1^{-2}\lambda_2^{-2} - 3\right)\right]^n - 1\right\} \quad (1)$$

where $\mu_0$ is the initial shear modulus, $\lambda_i (i=1,2)$ are principal stretches and the parameters $n$ and $b$ control the nonlinear response of the material. To achieve distinct nonlinear responses, we regulate $n$ and $b$ to obtain models exhibiting strain stiffening ($n=1.643$, $b=13.25$), large strain linear elasticity ($n=1.125$, $b=56.62$ obtained by fitting the Hooke's law, linear relation holds for $\lambda < 6$), and softening ($n=1$, $b=1$) in terms of nominal stress-stretch ratio, as depicted in Fig. 1(b). The dynamic fracture of these three materials is simulated using a well-established dynamic PFM [29]. In the PFM representation, sharp cracks are regularized by a scalar field $\phi \in [0,1]$, which smoothly transitions from $\phi=0$ (intact state) to $\phi=1$ (fully damaged).

Establishing the crack stability phase diagram often touches upon nonlinear length scale $\ell_n$ and crack driving force $G$ (i.e., the stored elastic energy per unit length). Taking cues from [24,35], we defined the normalized nonlinear scale

$$\ell_n / \xi = A \cdot \Gamma_0 / \mu_0 / \xi = A \cdot \ell_0 / \xi \tag{2}$$

and the normalized crack driving force

$$G / \Gamma_0 = W_m(\lambda) h_0 / \Gamma_0 = \tilde{W}_m(\lambda) h_0 / \ell_0 \tag{3}$$

Here, $\Gamma_0$ is the fracture energy, and $W_m(\lambda) = \mu_0 \tilde{W}_m(\lambda)$ represents the mean strain energy density of the intact material, hinging solely on the pre-strained state. The coefficient $A$ remains uncertain, typically with $A \gg 1$ for nonlinear elastic materials [25]. Considering the linear correlation, we define $\ell_0 = \Gamma_0 / \mu_0$ to characterize the variations in nonlinear scales. The phase-field regularization parameter $\xi$ [36,37], also termed the dissipative scale [38], serves as the normalized length unit. Then, we performed extensive simulations for the three materials in broad $G/\Gamma_0 - \ell_0/\xi$ parameter space. The results reveal that all crack morphologies can be classified into three distinct categories: stable (including cases where crack initiation is not possible), oscillating, and branching cracks. Fig. 2(a) demonstrates representative snapshots of these crack patterns (see supplementary movies [31]).

Based on the variations in crack morphologies, we further constructed the stability phase diagrams for three different materials (see Fig. 1(b)) in terms of the parameters $G/\Gamma_0$ and $\ell_0/\xi$, as showcased in Figs. 2(b)-(d). The obtained phase diagrams all consist of stable, oscillating, and branching phases, along with a forbidden region (vacant zone). We would like to emphasize that the pre-strain in the forbidden region exceeds the material's load-bearing capacity (the stored elastic energy density surpasses a critical threshold $W_*$), leading to intact material damage. In the available parameter space, the crack phase diagram of strain-stiffening materials exhibits an intriguing feature [Fig. 2(b)], where the oscillating and branching phases gradually vanish with the increase of $\ell_0/\xi$, eventually being completely occupied by the stable phase. For large-strain linear elastic materials [Fig. 2(c)], in contrast, the oscillating crack phase not only persists but also expands as $\ell_0/\xi$ raises. While in the case of strain-softening materials, the oscillating crack phase also persists with increasing $\ell_0/\xi$, it is confined to a narrow band [Fig. 2(d)]. These three phase diagrams suggest that the triggering conditions for oscillating cracks in materials exhibiting nonlinear softening or stiffening are highly demanding. In strain-softening materials, cracks tend to branch, whereas in strain-stiffening materials, stable cracks dominate. Remarkably, crack oscillation is more prone to arise in materials exhibiting the trait to sustain linear elasticity at large strains. This landscape elucidates why crack oscillation is seldom observed in fracture experiments of soft materials [18,19,39], except for brittle gels [13,14].

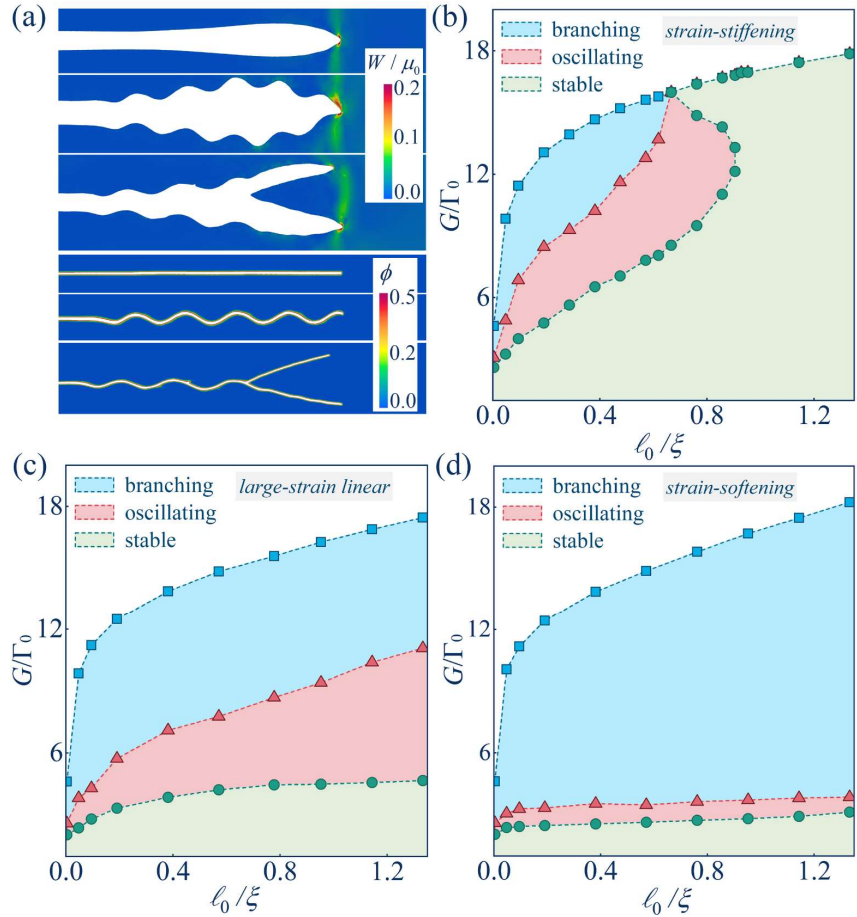

**FIG.2.** Effect of nonlinearity on crack dynamics. (a) Three distinct crack morphologies observed in simulations: stable, oscillating, and branching cracks. The rendered colors correspond to the normalized strain energy density $W/\mu_0$ in the deformed (upper) and phase field $\phi$ in the undeformed (lower) configurations. (b), (c) and (d) represent the crack stability phase diagrams of stiffening, large strain linear, and softening materials, respectively, in the $G/\Gamma_0 - \ell_0/\xi$ plane.

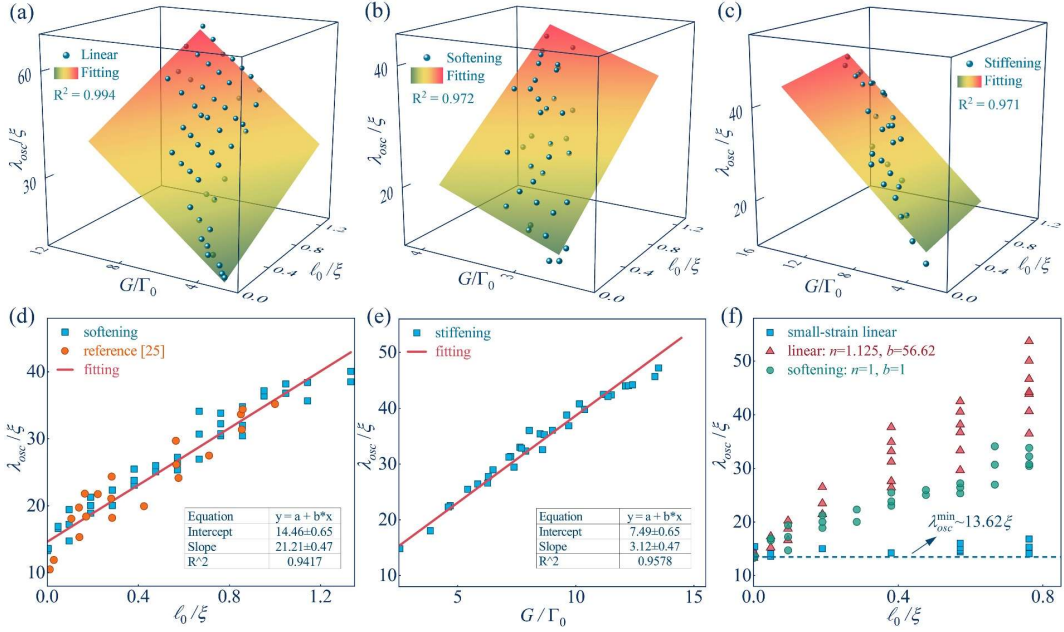

**FIG. 3.** Normalized oscillation crack wavelength $\lambda_{osc}/\xi$ as a function of variable pair $(G/\Gamma_0, \ell_0/\xi)$ for large strain linear elasticity (a), softening (b), and stiffening materials (c). The color of the fitted plane is rendered by the $\lambda_{osc}/\xi$ value. (d) $\lambda_{osc}/\xi$ vs $\ell_0/\xi$ for strain-softening materials (green data points extracted from Fig. 2(a) in [25]). (e) $\lambda_{osc}/\xi$ vs $G/\Gamma_0$ for strain-stiffening materials ($R^2 = 0.9578$). (f) Comparison of $\lambda_{osc}/\xi$ vs $\ell_0/\xi$ for small-strain linear elasticity, large-strain linear elasticity, and softening materials (each data point corresponds to a different value of $G/\Gamma_0$).

Concerning the three crack morphologies depicted in Fig. 2(a), the most appealing among them is the spontaneous crack oscillation characterized by a finite wavelength $\lambda_{osc}$. To explore the scaling law of $\lambda_{osc}$, we first measured $\lambda_{osc}$ in large-strain linear elastic materials existing a significant crack oscillating phase. As shown in Fig. 3(a), the normalized oscillation wavelength $\lambda_{osc}/\xi$ exhibits a bilinear function with respect to $\ell_0/\xi$ and $G/\Gamma_0$, i.e., $\lambda_{osc}/\xi = \alpha G/\Gamma_0 + \tilde{\beta}\ell_0/\xi + \delta$. The fitting parameters $\alpha = 4.57 \pm 0.13$ and $\tilde{\beta} = 15.85 \pm 0.54$ (with negligible $\delta \approx 0$) yield a high coefficient of determination with $R^2 = 0.994$, signifying a strong linear correlation. In the same manner, we measured $\lambda_{osc}$ of the other two materials, and fitted the obtained data with the bilinear function, as plotted in Figs. 3(b) and (c). For strain-softening materials [see Fig. 3(b)], the coefficients were identified as $\alpha = 5.75 \pm 0.27$ and $\tilde{\beta} = 15.48 \pm 1.23$ ($\delta \approx 0$), bearing a close resemblance to those of large-strain linear elastic materials. The calculated $R^2$ value of 0.972 suggests that $\lambda_{osc}$ remains a bilinear function of $(G/\Gamma_0, \ell_0/\xi)$ pair. However, we remark that previous studies on strain-softening gels only stated that $\lambda_{osc} \propto \ell_0$ without exploring the correlation between $\lambda_{osc}$ and $G/\Gamma_0$ [14,24,25]. To elucidate this issue, let us revisit the phase diagram shown in Fig. 2(d) for strain-softening materials. The variation of $G/\Gamma_0$ within the oscillation phase is minor, allowing it to be treated as a constant, thereby reducing the bilinear function to a simple linear relation: $\lambda_{osc}/\xi = \tilde{\beta}\ell_0/\xi + \delta$. With this in mind, we plot $\lambda_{osc}/\xi$ vs $\ell_0/\xi$ for strain-softening materials in Fig. 3(d). The subsequent linear regression evinces $\lambda_{osc} \propto \ell_0$, i.e., $\lambda_{osc} \propto \ell_n = A\ell_0$ as unveiled in experiments [14]. The fitted slope $\tilde{\beta} \approx 21.21$ and intercept $\delta \approx 14.63$ (with the minimum $\lambda_{osc}^{min} \approx 13.62\xi$) closely match the reported in [25].

As for strain-stiffening materials [see Fig. 3(c)], the fitting parameters are $\alpha = 2.67 \pm 0.21$, $\tilde{\beta} = 2.76 \pm 1.06$ and $\delta = 10.28 \pm 1.1$. Given the relative magnitudes of $\ell_0$ and $G/\Gamma_0$, a small $\tilde{\beta}$ implies that the contribution of $\ell_0$ to $\lambda_{osc}$ is minor, thus resulting in a simplified $\lambda_{osc}/\xi - G/\Gamma_0$ relation, as shown in Fig. 3(e). Note that, the evolution of $\lambda_{osc}$ in strain-stiffening materials exhibits a pronounced contrast compared to the other two materials. We speculate that this disparity stems from the strong nonlinearity induced by severe strain stiffening (see Fig.1(b)), which is reflected in $G/\Gamma_0$, while the weak nonlinearity characterized by $\ell_0$ takes a secondary role [27].

Figs. 3(a)-(d) also imply the existence of an intrinsic minimum scale $\lambda_{osc}^{min}$ for oscillation wavelengths as the elastic nonlinearity vanishes ($\ell_n = A\ell_0 \to 0$), in line with previous studies [28]. We then conducted a series of simulations in the small-strain linear elastic framework ($\ell_n = A\ell_0 = 0$, with $A=0$) to validate this finding. The outcomes strongly support the presence of an inherent scale $\lambda_{osc}^{min} \approx 13.62\xi$ [ see Fig.3(f)]. And besides, in the absence of any nonlinearity, we observed that $\lambda_{osc}$ exhibited no correlation with $\ell_0$ and was significantly smaller compared to nonlinear materials. This result verified that nonlinearity is not a prerequisite for the onset of supercritical crack oscillations, although it does serve to amplify the instability.

Since nonlinear response can regulate the dynamic patterns of fast-moving cracks, an ensuing issue emerges: how does it affect the crack speed, especially the onset speed of oscillation instabilities? To answer this, we measured the magnitude of normalized crack oscillation onset speed ($v_{onset}/c_s$) for the three different materials at various $\lambda$, as depicted in Fig. 4(a) [see Supplemental Material [31] for details]. For the softening materials, $v_{onset}$ is almost constant at about $0.92c_s$, in excellent agreement with [13,14]. Conversely, the dashed lines in Fig. 4(a), serving as visual references suggest a striking rule in the linear and stiffening materials, that is $v_{onset} \propto \lambda$, but with different slopes. Under identical macroscopic $\lambda$, the variations in $v_{onset}$ among the three materials hint at the possibility of crack dynamics being governed by local attributes. The crack tip zone, where strain is significantly amplified, comes to mind first. Enlightened by previous findings [18,20,28], we speculate that this phenomenon is related to the characteristic shear wave speed ($\tilde{c}_s$) in the vicinity of the crack tip, where $\tilde{c}_s = \sqrt{\mu_{non}/\rho}$. The nonlinear shear modulus $\mu_{non}$ for the adopted GNH model can be evaluated by [19,40]

$$\mu_{non} = \mu_0 \left[1 + \frac{b}{n}\left(\lambda_1^2 + \lambda_2^2 + \lambda_1^{-2}\lambda_2^{-2} - 3\right)\right]^{n-1} \quad (4)$$

For simplicity, the maximum $\mu_{non}$ near the crack tip at the onset of oscillation, termed $\mu_{non}^{tip}$, is extracted to estimate $\tilde{c}_s$.

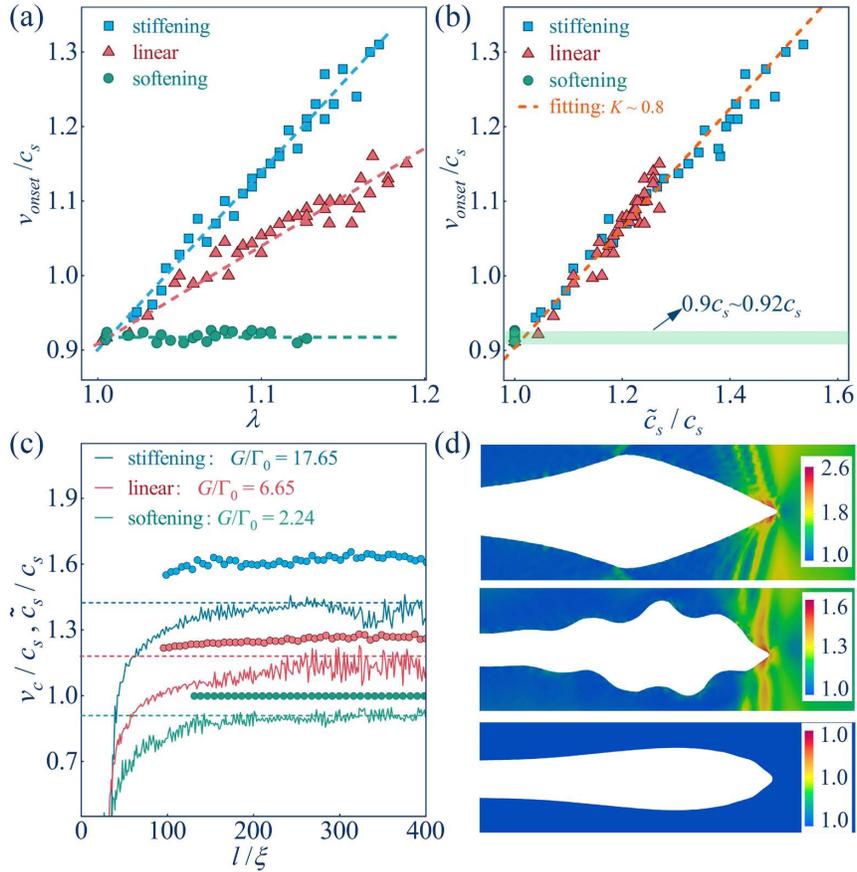

FIG. 4. (a) Comparison of $v_{onset}/c_s$ for three different materials at various $\lambda$. (b) Scatter plots of $v_{onset}/c_s$ vs $\tilde{c}_s/c_s$ for three different materials exhibiting approximate linear correlation. (c) Evolution of normalized crack speed $v_c/c_s$ and $\tilde{c}_s/c_s$ (scatter points) with the crack length $l$ (scaled by $\xi$) in $x_1$ direction for supershear fracture in stiffening materials, supercritical oscillations in the linear elastic regime, and sub-Rayleigh fracture modes in softening materials at the same $\ell_0 = 1.33$. Dashed lines indicate the limiting crack speed. (d) Snapshots of crack patterns corresponding to different fracture modes in (c) (from top to bottom). The color map represents the normalized nonlinear shear modulus $\mu_{non} / \mu_0$.

We then plotted $v_{onset} / c_s$ versus $\tilde{c}_s / c_s$ for the three materials, as presented in Fig. 4(b). Remarkably, the $v_{onset} / c_s$ values for the three distinct materials collapse onto a single linear trend, suggesting a universal nature of such crack oscillations that $v_{onset} / \tilde{c}_s \cong K$ holds for diverse materials. A compelling piece of evidence for this relation comes from fracture experiments on brittle gels described by NH model [14], where $v_{onset} \cong 0.9\tilde{c}_s = 0.9c_s$. In terms of the nominal stress-strain curve, the NH model exhibits softening behavior, yet Eq. 4 reveals that its $\mu_{non}$ remains constant, leading to $\tilde{c}_s = c_s$. In contrast, nominal large-strain linear elastic materials exhibit an escalating $\mu_{non}$ with $\lambda$ (see Fig. 4(d)). Besides, we also note that the linear $v_{onset} / c_s$ - $\tilde{c}_s / c_s$ relation yields a global slope of $K \cong 0.8$ [see Fig. 4(b)], slightly lower than the measured $K \cong 0.9$ in brittle gels [14]. This deviation may be attributed to the overestimation of $\tilde{c}_s$ resulting from the utilization of $\mu_{non}^{tip}$ (see Supplemental Material [31]).

Another intriguing phenomenon is the smooth transition from sub-Rayleigh to supershear fracture, accompanied by the alteration in crack morphologies. Figs. 4(c) portrays the normalized crack speed ($v_c / c_s$) evolutions in three representative scenarios: supershear fracture in strain-stiffening materials, crack oscillations in the linear elastic regime, and sub-Rayleigh straight cracks in the context of strain softening. The corresponding crack morphologies (with color indicating $\mu_{non} / \mu_0$) are presented in Fig. 4(d). The supershear crack exhibits a 'wedge-like' shape [18,41-43], distinct from the 'tadpole-like' morphology of the sub-Rayleigh crack [3,23]. Remarkably, this feature has also been observed in recent experiments on large-strain dynamic fracture of brittle gels [8]. For quantitative comparison, we further measured the asymptotic crack speeds (the dash line in Fig. 4(c)) for the three cases to be roughly 1.42$c_s$, 1.18$c_s$, and 0.89$c_s$, while the calculated $\tilde{c}_s$ using the aforementioned strategy yielded values of approximately 1.61$c_s$, 1.28$c_s$, and $c_s$, respectively (see the scatter points in Fig. 4(c)). Such results signify that the crack speed in nonlinear materials is still limited by $\tilde{c}_s$ near the crack tip. Moreover, the existing evidence strongly suggests that Mode-I supershear fracture arises from strain stiffening at the crack tip, whereas it is absent in homogeneous soft materials exhibiting strain-softening characteristics (introducing a weak layer could potentially lead to this case [8]).

At this point, we have gained an in-depth understanding of how nonlinearity regulates crack dynamics. The versatile GNH model allows for manipulating the nonlinearity of materials. Utilizing this model, we have constructed phase diagrams depicting crack stability for various nonlinear materials in the extensive $\ell_0/\xi - G/\Gamma_0$ plane. In strain-softening materials, crack branching dominates, whereas in strain-stiffening materials, straight cracks prevail. Of particular note is the large-strain linear elastic material, wherein crack oscillation is easily triggered. The three distinct phase diagrams highlight the demanding nature of triggering crack oscillation in nonlinear soft materials. A conjecture arising from these findings is that the crack oscillation is more difficult to occur in materials exhibiting more pronounced strain stiffening or softening. Besides, our simulations also suggest a novel scaling law that demonstrates $\lambda_{osc}$ as a bilinear function of $G/\Gamma_0$ and $\ell_0/\xi$. For strain-stiffening or softening materials, this relation is reasonably approximated as a linear function of either $G/\Gamma_0$ or $\ell_0/\xi$. Based on small strain linear elasticity, we further verify the existence of an inherent oscillation wavelength $\lambda_{osc}^{\min} \approx 13.62\xi$ as $\ell_n = A\ell_0$ vanishes.

Apart from regulating crack morphology, nonlinearity also markedly alters the crack speed. When the nonlinearity response is adjusted from softening to stiffening, the crack velocity shifts from sub-Rayleigh to supershear, accompanied by a transition in the crack profile from smooth to sharp. By introducing the characteristic shear wave speed, $\tilde{c}_s$, we ascertain that the underlying cause of supershear fracture lies in the strain-stiffening-induced increase of $\tilde{c}_s$, enabling faster crack propagation while still limited by $\tilde{c}_s$. More interestingly, we discovered that the crack oscillation velocity, $v_{onset}$, exhibits a nearly identical linear relation with $\tilde{c}_s$ across three distinct materials. This scaling law unveils the nature of crack oscillation as a universally present instability closely tied to the wave speed.

To sum up, our findings unveil the universal laws of nonlinearity in regulating fracture dynamics, significantly enriching the theoretical connotation of nonlinear fracture physics and mechanics. However, we also remark that recent experiments suggest the existence of a novel supershear fracture paradigm, excited at critical strains [8], which cannot be accounted for within the current theoretical framework [43]. Future work will delve deeper into the unified mechanisms governing the transition of super-shear cracks.

## Acknowledgments

The authors acknowledge the support from JSPS International Research Fellow in Japan (Grant No. 23KF0002). J.P.G. thanks the JSPS KAKENHI (No. 22H04968, 22K21342). F. T. thanks the National Natural Science Foundation of China (Grant No. 12102420).

**Supplemental Material for:**

**Nonlinearity tunes crack dynamics in soft materials**

Fucheng Tian[a], Jian Ping Gong [a, b]

[a] *Faculty of Advanced Life Science, Hokkaido University, Sapporo 001-0021, Japan;*
[b] *Institute for Chemical Reaction Design and Discovery (WPI-ICReDD), Hokkaido University, Sapporo 001-0021, Japan*

Here we present the theoretical aspects of the dynamic fracture phase-field model, details on the simulation setup, post-processing numerical techniques, and some supplementary results.

### S-1. DYNAMIC PHASE FIELD MODEL

The phase field model (PFM) utilized in this study is rooted in the classical Griffith's theory [1,2], which has been widely recognized in fracture modeling [3,4]. This approach avoids the cumbersome tracking discontinuous crack surfaces and seamlessly incorporates into the computational framework of continuum mechanics. To simulate high-speed fracture in soft materials, we employ a recently developed dynamic version of PFM with no attenuation of wave speed [5]. The strong form governing equations involving two unknowns, the displacement field $\mathbf{u}$ and the phase field $\phi$, derived from the non-conservative Lagrangian formulation (interested readers are referred to [5] for more details on the concrete derivation), are expressed as follows

$$\begin{cases} \rho f(\phi)\ddot{\mathbf{u}} + \rho \frac{\partial f(\phi)}{\partial t}\dot{\mathbf{u}} = \nabla_X \cdot \mathbf{P} + \overline{\mathbf{b}}_0 \\ \frac{1}{4c_w}(\frac{\Gamma_0}{\xi} - 2\Gamma_0 \xi \nabla^2 \phi) + \eta\dot{\phi} - \frac{\rho}{2}\frac{\partial f(\phi)}{\partial \phi}\dot{\mathbf{u}} \cdot \dot{\mathbf{u}} = 2(1-\phi)W \\ \mathbf{P} \cdot \mathbf{n} = \overline{\mathbf{t}}_0 \qquad \text{at } \partial\Omega_0 \\ \nabla_X \phi \cdot \mathbf{n} = 0 \qquad \text{at } \partial\Omega_0 \end{cases} \quad (1)$$

Herein, $\overline{\mathbf{b}}_0$ and $\overline{\mathbf{t}}_0$ are the body and traction force vectors, respectively (the body force can be neglected). $\mathbf{n}$ represents the outward norm $f(\phi) = (1-\phi)^2 + k$ ($0 < k \ll 1$) al direction vector of boundary $\partial\Omega_0$. The degradation function commonly takes the form as [1,2]. The viscosity coefficients $\eta$, which controls the rate of energy dissipation, adheres to the relation $\eta = \beta \dfrac{\Gamma_0}{2c_w c_s}$ [5], where $\Gamma_0$ is the fracture energy, and $\beta$ is a dimensionless parameter, set as $\beta = 0.1$ [6]. For small $\beta$, the fracture energy is almost independent of crack speed (see [6] for a discussion on the value of $\beta$). $c_s$ is the shear wave speed, defined by $c_s = \sqrt{\mu_0/\rho}$, where $\mu_0$ denotes the shear modulus and $\rho$ is the density of the materials. For the used PFM [2], we have

$$c_w = \int_0^1 \sqrt{\phi}\, \mathrm{d}\phi = \dfrac{2}{3} \tag{2}$$

As stated in main text, we consider the following incompressible plane-stress generalized neo-Hookean (GNH) model, thus the stored strain energy density $W$ (see Eq. 1) is given by

$$W = \dfrac{\mu_0}{2b}\left\{\left[1+\dfrac{b}{n}\left(\lambda_1^2 + \lambda_2^2 + \lambda_1^{-2}\lambda_2^{-2} - 3\right)\right]^n - 1\right\} \tag{3}$$

allowing for flexible tuning of the material's nonlinear response. Here, $\lambda_i (i=1,2)$ are principal stretches. To streamline program implementation, we reformulate Eq.3 as

$$W = \dfrac{\mu_0}{2b}\left\{\left[1+\dfrac{b}{n}\left(\mathrm{tr}(C_{ij}) + C_{33} - 3\right)\right]^n - 1\right\} \tag{4}$$

where $C_{ij}(i,j=1,2)$ are the in-plane components of the right Cauchy-Green tensor $\mathbf{C}$, namely

$$\mathbf{C} = \begin{bmatrix} C_{ij} & 0 \\ 0 & C_{33} \end{bmatrix} \tag{5}$$

In the incompressible case, we have the out-of-plane component $C_{33} = J_0^{-2}$, in which $J_0 = \det(F_{ij})$ is the in-plane Jacobian determinant. $F_{ij} = \delta_{ij} + \partial_j u_i$ are the in-plane components of deformation gradient tensor ($\delta_{ij}$ represent the components of second-order unity tensor).

Using Eq. 4, we derived the second Piola-Kirchhoff stress (PK2) as

$$S_{ij} = 2f(\phi)\dfrac{\partial W}{\partial C_{ij}} = \mu_0 f(\phi)\left[1+\dfrac{b}{n}\left(\mathrm{tr}(C_{ij}) + J_0^{-2} - 3\right)\right]^{n-1}\left[\delta_{ij} - J_0^{-2}C_{ij}^{-1}\right]. \tag{6}$$

Thus, the components of the nominal stress $\mathbf{P}$ in Eq. 1 can be given by

$$P_{ij} = S_{ij}F_{ij}^T = \mu_0 f(\phi)\left[1+\dfrac{b}{n}\left(\mathrm{tr}(C_{ij}) + J_0^{-2} - 3\right)\right]^{n-1}\left[F_{ij}^T - J_0^{-2}F_{ij}^{-1}\right]. \tag{7}$$

In the quasi-static tensile simulations, we also derived the components $\mathbb{C}_{ijkl}(i,j,k,l=1,2)$ of fourth-order elastic tensor

$$\mathbb{C}_{ijkl} = 2\frac{\partial S_{ij}}{\partial C_{kl}} = 2\mu_0 f(\phi) J_0^{-2}\left[1+\frac{b}{n}\left(\text{tr}(C_{ij})+J_0^{-2}-3\right)\right]^{n-1}\left[C_{ij}\odot C_{kl}+C_{ij}\otimes C_{kl}\right]+$$
$$2\mu_0 f(\phi)(n-1)\frac{b}{n}\left[1+\frac{b}{n}\left(\text{tr}(C_{ij})+J_0^{-2}-3\right)\right]^{n-2}\left[\delta_{ij}-J_0^{-2}C_{ij}^{-1}\right]\otimes\left[\delta_{kl}-J_0^{-2}C_{kl}^{-1}\right] \quad (8)$$

for programming purposes (tensor differentiation can be referred to [7]). However, $\mathbb{C}_{ijkl}$ is unnecessary for the explicit dynamic fracture solving procedure [5].

To obtain universal results, we here introduce the dimensionless version of Eq. 1. The non-dimensionalization rules are consistent with [6,8,9], where $\xi$ serves as the length unit and $\tau = \beta\xi/c_s$ acts as the time unit. By substituting the density $\rho = \mu_0/c_s^2$, viscosity $\eta = \beta\Gamma_0/2c_w c_s$, $\ell_0 = \Gamma_0/\mu_0$ and $\tilde{W} = W/\mu_0$, Eq. 1 can be recast as

$$\begin{cases} f(\phi)\ddot{\tilde{\mathbf{u}}} + \frac{\partial f(\phi)}{\partial \tilde{t}}\dot{\tilde{\mathbf{u}}} = \beta^2 \nabla_{\tilde{X}} \cdot \tilde{\mathbf{P}} \\ \frac{1}{4c_w}(\frac{\ell_0}{\xi} - 2\frac{\ell_0}{\xi}\nabla^2\phi) + \frac{\ell_0}{2\xi c_w}\frac{\partial \phi}{\partial \tilde{t}} - \frac{1}{2\beta^2}\frac{\partial f(\phi)}{\partial \phi}\dot{\tilde{\mathbf{u}}}\cdot\dot{\tilde{\mathbf{u}}} = 2(1-\phi)\tilde{W} \\ \tilde{\mathbf{P}}\cdot\mathbf{n} = \overline{\mathbf{t}}_0 \quad \text{at } \partial\Omega_0 \\ \nabla_X\phi\cdot\mathbf{n} = 0 \quad \text{at } \partial\Omega_0 \end{cases} \quad (9)$$

Here, the superscript '~' denotes the dimensionless form of the physical quantities, i.e., the displacement field $\tilde{\mathbf{u}} = \mathbf{u}/\xi$, the nominal stress $\tilde{\mathbf{P}} = \mathbf{P}/\mu_0$, and the time scale $\tilde{t} = t/\tau$. Clearly, for a given $\beta$, the solution of Eq. 9 depends solely on material-related parameter $\ell_0/\xi$ and $\tilde{W}$ related to the applied boundary condition (pre-strain). This also lays the theoretical groundwork for constructing the phase diagram (see Fig. 2) in the main text. Unlike $\ell_0/\xi$, however, $\tilde{W}$ is not a single value but rather a distribution dependent on the strain field. For simplicity, we require a concise scalar to characterize it, and the crack driving force $G$ is the optimal candidate for this purpose. We normalize $G$ by fracture energy $\Gamma_0$ to render it dimensionless, such that

$$G/\Gamma_0 = W_m(\lambda)h_0/\Gamma_0 = \tilde{W}_m(\lambda)h_0/\underbrace{(\Gamma_0/\mu_0)}_{\ell_0} = \tilde{W}_m(\lambda)h_0/\ell_0 \quad (10)$$

Here, we defined the mean strain energy density in the intact sample

$$W_m(\lambda) = \mu_0\tilde{W}_m(\lambda) = \frac{1}{V}\int W_{intact}(\lambda)dV \quad (11)$$

where, $h_0$ and $V$ respectively denote the height and volume (area for 2D) of the computational domain (see Fig. S2). $W_{intact}(\lambda)$ is the distribution of strain energy density in the intact sample. From Eq. 10, we find that $G/\Gamma_0$ does not depend on the specific value of $\Gamma_0$, but scales linearly with $h_0$ for a given $\ell_0$.

**Remark.** For an infinite sample length, the stored strain energy density $W(\lambda)$ of the intact sample can be considered independent of position. However, in our simulations, the length of the sample is finite ($w_0$), so we accurately account for the position dependence of $W(\lambda)$ due to edge effects.

## S-2. DETAILS ON THE SIMULATION SETUP

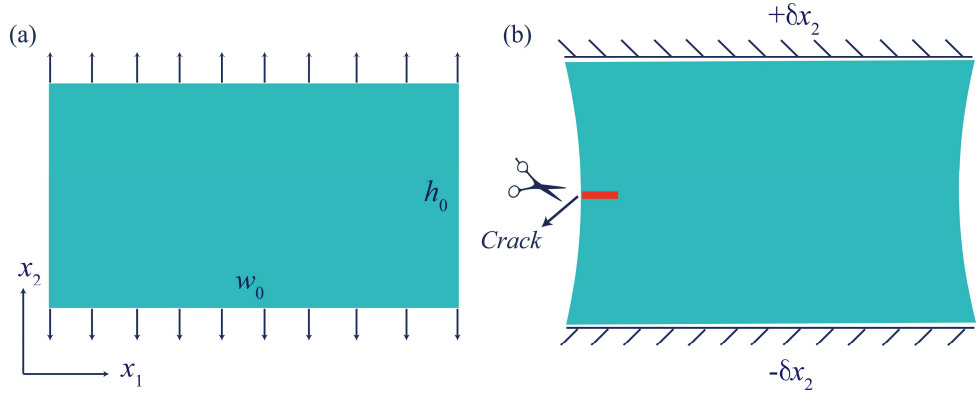

Fig. S1. Illustration of the geometric configuration ($w_0 \times h_0 = 480 \times 240 \xi^2$) and applied boundary conditions in the initial (a) and pre-strained states (b).

The dynamic fracture modeling in this study follows a 2D pre-strained fracture configuration that closely resemble experimental setups [10,11]. The initial geometric model is a pristine rectangular sheet with $480\xi$ in width ($w_0$), $240\xi$ in height ($h_0$), as illustrated in Fig. S1(a). We first impose uniform vertical displacement loads $\pm\delta x_2$ on the top and bottom edges of the specimen under a quasi-static framework until the desired strain state is attained. Maintaining this strain state, dynamic fracture is triggered by a seed crack with length of $20\xi$ at the middle height of the left edge (see Fig. S1(b)), created by setting the phase field value $\phi=1$. As $\xi$ serves as the normalized unit length, its specific value does not affect the inherent self-similarity of the solutions to the dimensionless Eq. 7. From this perspective, scaling the geometric model may result in specific numerical differences, but the similarity of the solutions (physical laws) remains unchanged. To ensure comparability with previous reports [8,9], we set the phase field characteristic scale as $\xi=153$ μm. In a nutshell, we employed an adaptive finite element method for spatial discretization of the solution domain, with an effective element size of $h_e = \xi/4$ [12]. An explicit velocity-Verlet algorithm was used for time discretization, with the time step of $\Delta t = 0.04\tau < h_e/c_s$ [12].

## S-3. NONLINEAR MATERIAL MODEL

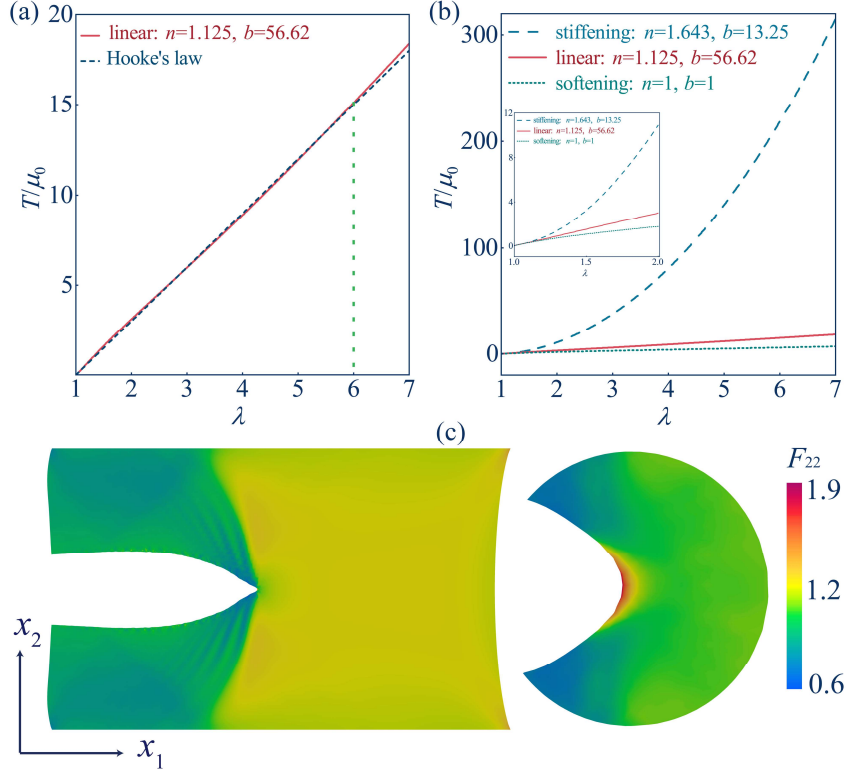

FIG. S2. (a) Normalized nominal stress ($T/\mu_0$) -stretch ratio ($\lambda$) curves of the large-strain linear elastic GNH model fitted with Hooke's law (approximate linear relation holds for $\lambda < 6$). (b) $T/\mu_0$ versus $\lambda$ for the three materials within a wide range of $\lambda$. (c) Maximum tensile scenario involved in the large-strain linear elastic material (macroscopic stretch ratio $\lambda \approx 1.22$). The locally maximum stretch ratio near the crack tip, evaluated by the deformation gradient component $F_{22}$ (stretch ratio along $x_2$ direction), is approximately $\lambda_{max} \approx 1.94$.

In the main text, we focus on three distinct nonlinear materials described by the GNH model, namely strain stiffening ($n = 1.643$, $b = 13.25$), large strain linear elasticity ($n = 1.125$, $b = 56.62$, linear relation holds for $\lambda < 6$), and softening ($n = 1$, $b = 1$) in terms of nominal stress-stretch ratio. Fig 1(b) in the main text is plotted based on the uniaxial tension formula of the GNH model

$$T = \mu_0 \left\{ 1 + \frac{b}{n}\left(\lambda^2 + \frac{2}{\lambda} - 3\right) \right\}^{n-1} \left(\lambda - \frac{1}{\lambda^2}\right) \qquad (12)$$

where $T$ represents the nominal uniaxial stress and $\lambda$ is the uniaxial stretch ratio. The so-called softening model referred to here is the classical neo-Hookean (NH) model, which possesses an invariant shear modulus but exhibits softening behavior in the nominal stress. The strain stiffening behavior is similar to the mechanical response described by the Saint-Venant-Kirchhoff (SVK) model [7]. It should be note that the parameters ($n$ =1.125, $b$ =56.62) for large-strain linear elasticity were obtained through fitting the Hooke's law ($T = 3\mu_0(\lambda - 1)$), and the linear relation holds approximately for $\lambda < 6$, as shown in Fig. S2(a). While this model exhibits approximate linearity in nominal stress, its shear modulus slightly increases with strain. From Eq. 1, it is evident that the nominal stress **P** directly influences the crack dynamics.

The complete mechanical responses of the three materials for $\lambda < 7$ are shown in Fig. S2(b). To highlight the disparities between the three materials, we only presented the data for $\lambda < 2$ in the main text (see the localized zoom-in view in Fig. S2 (b)). Besides, we also remark that singularities near the crack tip can significantly amplify the strains. We present the maximum tensile scenario involved in the fracture modeling of the large-strain linear elastic material, as depicted in Fig. S2(c). The zoomed-in view reveals that the reached maximum stretch ratio ($\lambda_{max}$) is approximately 1.94, corresponding to a macroscopic stretch ratio of about 1.22 (the strain amplification is approximately 4 to 5 times). Thus, the fitted range of large-strain linearity ($1 \leq \lambda \leq 6$) adequately covers all simulated scenarios in this work.

**S-4. DIVERSE CRACK MORPHOLOGIES**

Here, we showcase a range of observed crack morphologies from the reported experiment as supporting evidence for our simulations. In the dynamic fracture simulations of three distinct soft materials, we have observed diverse crack morphologies classified into three phases: stable (straight) cracks, oscillating cracks, and branching cracks (see Figs. S3-S5). We would like to emphasize that seed cracks do not propagate under relatively low crack driving forces, which is not a focus of the current issue and is directly classified as a stable state.

In Fig. S3(a), we present two distinct straight crack patterns observed in our simulations, i.e., the 'tadpole-shaped' sub-Rayleigh crack (top) with crack speed $v_c \approx 0.89 c_s$ in the strain-softening material and the 'wedge-shaped' supershear crack (bottom) with $v_c \approx 1.44 c_s$ in the strain-stiffening material. Both crack patterns have been captured in fracture experiments on brittle gels [13], as depicted in Fig. S3(b). We want to emphasize that, the emergence of supershear fracture in our simulations is attributed to the localized increase in wave speed near the crack tip caused by stiffening. However, in the experiments conducted by Wang et al. [13], supershear fracture can occur in homogeneous softening materials, which is astonishing and may reveal a brand new mechanism, provided that we can clarify the absence of strain stiffening near the crack tip (If a weak interface is introduced, the crack can attain supersonic speeds even in softening materials).

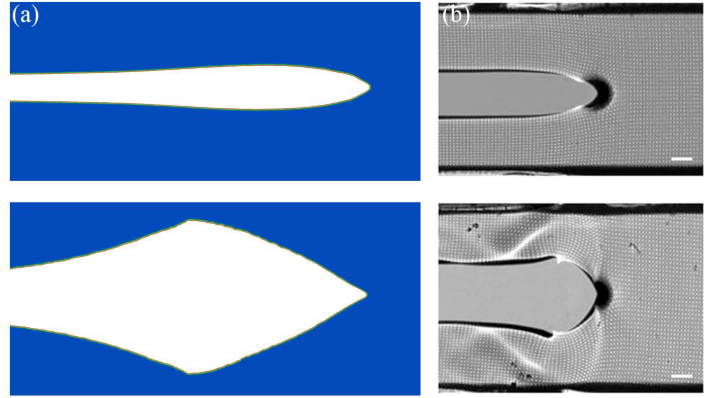

Fig. S3. (a) Two distinct straight crack patterns observed in our simulations. Top: Sub-Rayleigh straight crack in the strain-softening material at ($G/\Gamma_0$, $\ell_0/\xi$)=(2.24, 1.33); Bottom: supershear (bottom) straight crack in the strain-stiffening material at ($G/\Gamma_0$, $\ell_0/\xi$)=(17.65, 1.33). (b) Sub-Rayleigh (top) and supershear (bottom) straight cracks captured in the experiments of Wang et al [13].

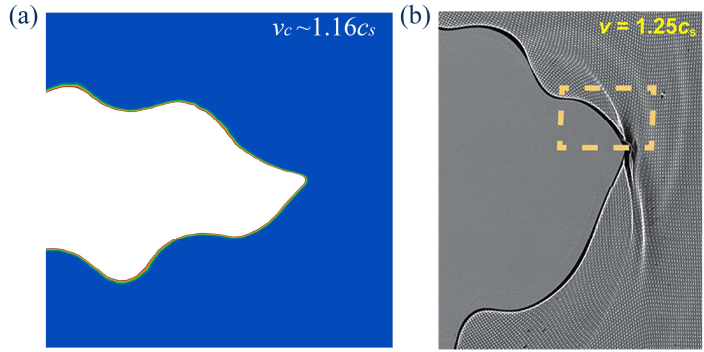

Fig. S4. Snapshots of crack oscillation instabilities captured in fracture simulations of large strain linear elastic materials at ($G/\Gamma_0$, $\ell_0/\xi$)=(6.65, 1.33) (a) and Wang's experiments (b) [13] in the supershear regime.

We have revealed that the spontaneous crack oscillations are a universal fracture instability closely tied to the wave speed. The onset velocity of these oscillations linearly scales with the characteristic wave speed $\tilde{c}_s$ near the crack tip. This mechanism enables the occurrence of crack oscillations in supershear regimes. Fig. S4(a) presents a snapshot of supershear crack oscillation in our simulations, with an instantaneous crack speed of approximately $v_c \sim 1.16 c_s$. Similar supershear oscillation has also been observed in recent experiments [13] (see Fig. S4(b)), with corresponding crack speed of $1.25 c_s$. Apart from crack oscillations, our simulations also captured diverse crack branching patterns, where crack branching can occur either before or after oscillations, as shown in Figs. S5(a, c, d). Analogous crack patterns have been reported in the fracture of brittle gels [6,9], as displayed in Fig. S5(b). These experimental observations strongly validate the reliability of our simulations.

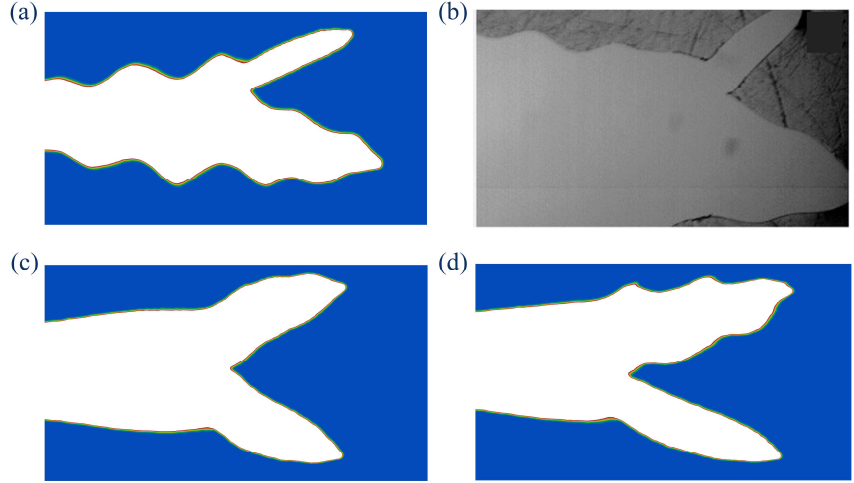

Fig. S5. Snapshot of crack branching observed in the fracture simulations of large strain linear elastic materials (a: $G/\Gamma_0 = 7.57$, c: $G/\Gamma_0 = 8.43$, d: $G/\Gamma_0 = 8.87$ for same $\ell_0/\xi = 0.57$) and experiments (b) [9]. Crack branching can occur independently or accompany path oscillations.

## S-5. MEASUREMENT OF OSCILLATION WAVELENGTH

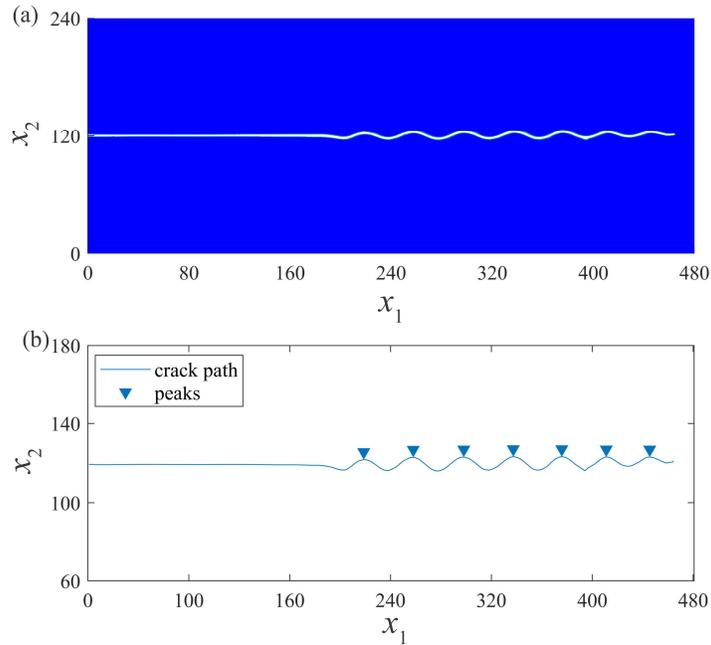

FIG. S6. Schematic diagram for measuring crack oscillation wavelength ($\ell_0/\xi = 0.57$ and $G/\Gamma_0 = 6.65$). (a) Crack patterns in the undeformed configuration. (b) Amplified view of the extracted crack path using MATLAB code.

In almost all simulations, we observed that the wavelength of crack oscillation exhibits slight fluctuations rather than remaining constant throughout the entire fracture process. Previous studies have used the first oscillation as a measurement of wavelength. However, we remark that the initial crack oscillation is typically incomplete (see Fig. S6), and utilizing it for wavelength evaluation will introduce large disturbances to the data. For this reason, we consider the average wavelength of all complete oscillations as a measurement of the wavelength $\lambda$ for a complete fracture scenario, such that $\lambda_{osc} = \frac{1}{n}\sum_{i=1}^{n}\lambda_{osc}^{i}$. To accurately identify all oscillation wavelengths, a MATLAB script was developed to locate the peaks of all complete oscillations, as demonstrated in Fig. S6, enabling the calculation of $\lambda_{osc}$.

## S-6. LIMITING CASE: VANISHING NONLINEAR SCALE

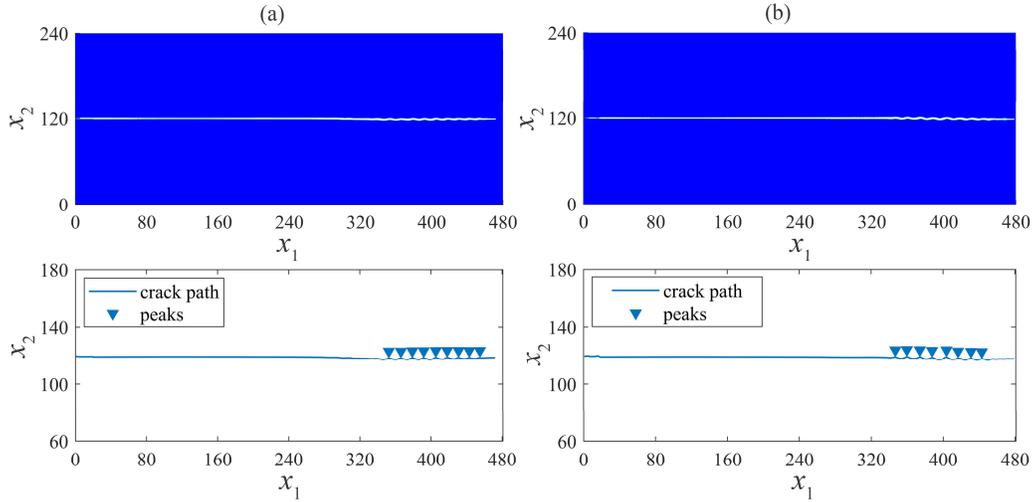

FIG. S7. (a) Crack oscillation pattern in large strain linear elastic materials ($n=1$, $b=1$) with parameters: $\ell_0/\xi \approx 1\times 10^{-3}$, $G/\Gamma_0 = 2.54$. $\ell_0/\xi \approx 1\times 10^{-3}$ (b) Crack oscillation pattern in small strain linear elastic materials with parameters: $\ell_0/\xi \approx 5\times 10^{-2}$, $G/\Gamma_0 = 2.43$.

For the employed GNH model, we reduced $\ell_0/\xi$ to the order of $O(10^{-3})$, and yet crack oscillations still occurred with a minimum wavelength of $\lambda_{osc}^{\min} = 13.62\xi$, as shown in Fig. S7(a). This result strongly suggests that crack oscillations can occur even for $\ell_0 \to 0$. However, this approach still cannot touch upon the limit case of $\ell_0 = 0$. To address this, we directly employ a material model that does not involve any geometric or material nonlinearity, that is, the linear elastic model under small strain assumption, with the strain energy density given by

$$W(\boldsymbol{\varepsilon}) = \mu_0 \left( \left[ tr(\boldsymbol{\varepsilon}) \right]^2 + tr(\boldsymbol{\varepsilon}^2) \right) \tag{13}$$

where $\boldsymbol{\varepsilon} = \frac{1}{2}(\nabla \mathbf{u} + \nabla \mathbf{u}^T)$ is the linear elastic strain tensor. This model does not incorporate any nonlinearity, implying that $\ell_n = A\ell_0 = 0$ holds true regardless of the values of $\ell_0 = \Gamma_0 / \mu_0$, i.e., $A=0$. One of the simulated patterns of crack oscillation is displayed in Fig. S7(b), exhibiting an oscillation wavelength of approximately $13.62\,\xi$. This phenomenon has also been observed in the simulations by Vasudevan et al.[6]. We then plot $\lambda_{osc}/\xi$ vs $\ell_0/\xi$ in the context of small-strain linear elastic materials (scatter points correspond to various $G/\Gamma_0$) in Fig. S8(a) (the data corresponds to the Fig. 3(f) in the main text). As observed, $\lambda_{osc}$ exhibits no correlation with $\ell_0$. However, once we replace the horizontal axis with $G/\Gamma_0$, the $\lambda_{osc}/\xi$ - $G/\Gamma_0$ data exhibit linear correlation, as shown in Fig. S8(b) (the orange line serves as a visual guide). This finding reinforces the conclusion in the main text that the crack oscillation wavelength is a bilinear function of both $\ell_0/\xi$ and $G/\Gamma_0$.

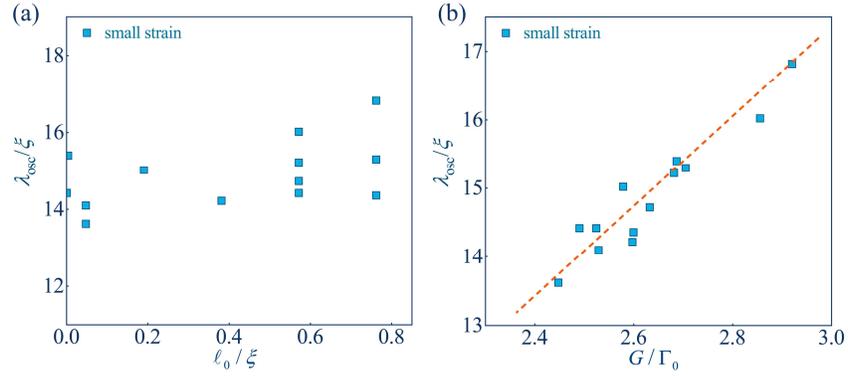

FIG. S8. (a) $\lambda_{osc}/\xi$ vs $\ell_0/\xi$ (a) and $\lambda_{osc}/\xi$ vs $G/\Gamma_0$ (b) for small-strain linear elastic materials. scatter points correspond to various $G/\Gamma_0$.

## S-7. CHARACTERISTIC SHEAR WAVE SPEED

The concept of characteristic shear wave speed ($\tilde{c}_s$) introduced here is enlightened by the notion of characteristic length scale for the energy flux near the crack tip proposed by Buehler et al [14]. Our idea is to consider the effective shear wave speed within the characteristic length scale as $\tilde{c}_s$. However, quantifying the characteristic length scale near the fast-moving crack tip remains a challenging task. For simplicity, we regard the maximum shear wave speed near the forefront of the crack dissipation zone as $\tilde{c}_s$. Fig. S9 demonstrates the determination of $\tilde{c}_s$ for a crack oscillation scenario ($\ell_0/\xi = 0.76$ and $G/\Gamma_0 = 6.65$). The visualization of crack opening profile is achieved by eliminating continuum with the level set of $\phi = 0.01$, which can be approximated as the forefront of the crack dissipative zone. We extracted the maximum $\mu_{non}$ near the crack tip, termed as $\mu_{non}^{tip}$, such that $\tilde{c}_s$ can be estimated by

$$\tilde{c}_s = \sqrt{\mu_{non}^{tip}/\rho} \qquad (14)$$

For the snapshot shown in Fig. S9, we get $\tilde{c}_s \approx 1.19 c_s$ by Eq. 15. This approach is expected to lead to an overestimation of $\tilde{c}_s$, but provides a more unified framework (subsequent Fig. S10 will demonstrate the robustness of this method). The current estimation of $\tilde{c}_s$ yields a slope of approximately 0.8 for the $v_{onset} - \tilde{c}_s$ relation, which is lower than the measured value of 0.9 in brittle gel fracture experiments [8,15,16]. It should be noted that $\tilde{c}_s$ is a distribution related to strain field, and this approach to estimating $\tilde{c}_s$ is evidently rudimentary (potentially overestimating). We believe a refined assessment of $\tilde{c}_s$ will approach the experimental results. There is still much work to be done on this topic in the future.

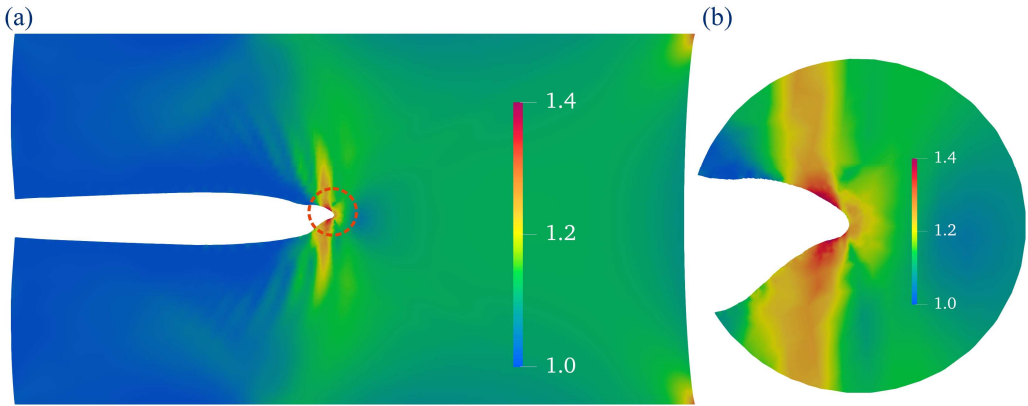

FIG. S9. Schematic diagram for measuring the characteristic shear wave velocity for a crack oscillation scenario ($\ell_0/\xi = 0.76$ and $G/\Gamma_0 = 6.65$). (a) Distribution of the nonlinear shear modulus ($\mu_{non}/\mu_0$). (b) A local zoom-in view near the crack tip for the region enclosed by red dashed circle in (a).

## S-8. MEASUREMENT OF CRACK SPEED

For fracture phase-field modeling, the exact position of the crack tip is not explicitly provided. Following prior work [5,8], we consider the level set $\phi = 0.5$ as the crack profile. The forefront of the profile is identified as the crack tip. Assuming that the crack tip propagates a distance of $\Delta x$ within a time step $\Delta t$, the magnitude of the crack velocity $v_c$ can be calculated by

$$v_c = \frac{\Delta x}{\Delta t} = \frac{\sqrt{\Delta x_1^2 + \Delta x_2^2}}{\Delta t} \tag{15}$$

Here, $\Delta x_1$ and $\Delta x_2$ are the crack propagation components in the $x_1$ and $x_2$ directions, respectively.

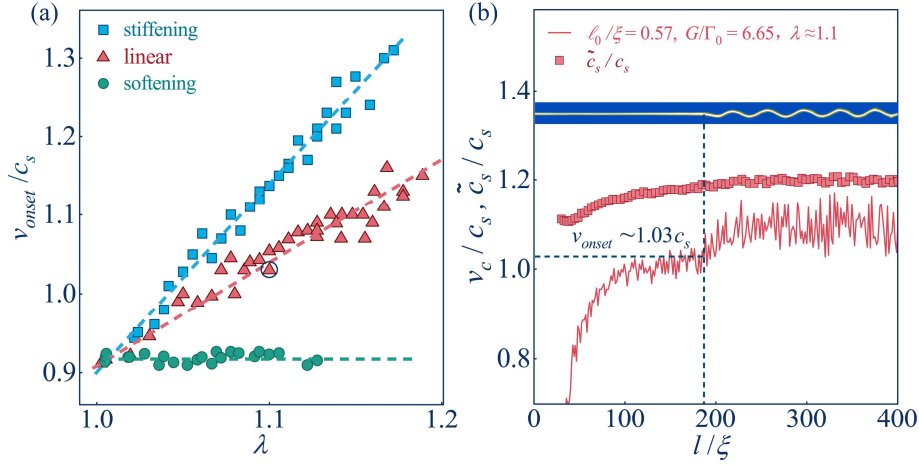

FIG. S10. (a) Comparison of $v_{onset}/c_s$ for three different materials at various $\lambda$ (see Fig. 4(a) in the main text). (b) An example for identifying the onset velocity ($v_{onset}$) of crack oscillation in large strain linear elastic materials with parameters: $\ell_0/\xi = 0.57$, $G/\Gamma_0 = 6.65$ (the circled point in Fig.S10(a)). The scatter points represent instantaneous values of $\tilde{c}_s/c_s$. For comparison, the crack morphology is also placed in this figure.

In the main text, the onset velocity ($v_{onset}$) is a characteristic quantity of interest. Here, we illustrate the determination of $v_{onset}$ using the oscillatory scenario depicted in Fig. S6 (large strain linear elastic materials with parameters: $\ell_0/\xi = 0.57$, $G/\Gamma_0 = 6.65$). For the sake of comparison, we have reproduced the Fig. 4(a) from the main text in Fig. S10(a). The crack morphology and the evolution of the normalized $v_c/c_s$ with the crack length $l$ (scaled by $\xi$) in $x_1$ direction, corresponding to the scenario depicted in Fig. S6, are presented in Fig. S10(b). As observed, $v_{onset}$ for this scenario is identified to be approximately $1.04c_s$, corresponding to the circular marker in Fig. S10(a). We also remark that after the occurrence of crack oscillations, the normalized $v_c/c_s$ experiences a significant uplift due to the non-zero velocity component in the $x_2$ direction. This trait has also been discovered in previous simulations and experiments [8,9,13]. Furthermore, we employed the approach described in S-7 section to calculate $\tilde{c}_s$, as illustrated by the scatter points in Fig. S10(b). After the occurrence of crack oscillation, the value of $\tilde{c}_s$ remains nearly constant at approximately 1.19$c_s$. This outcome is consistent with the result depicted in Fig. 4(c) of the main text, thereby re-corroborating the coherence of the approach presented in S-7.

**S-9. SUPPLEMENTAL MOVIES**

The supplementary four movies illustrate the evolution of four distinct crack morphologies. Brief descriptions of each movie are presented below.

Movie_1: The crack exhibits a sub-Rayleigh stable state with a limiting speed of $v_{\text{limit}} \approx 0.9 c_s$, the color of which is rendered by $\sigma_2 / \mu_0$ (see the first crack pattern (rendered by $W/\mu_0$) in Fig. 2(a)). The utilized GNH model parameters are $n=1$ and $b=1$, corresponding to a strain-softening material. The crack state is located in the crack stability phase diagram (Fig.2(d)) at ($G/\Gamma_0$, $\ell_0/\xi$)=(2.54, 0.57).

Movie_2: This movie depicts a scenario of crack oscillation instability in the large strain linear elastic material (see the second crack pattern in Fig. 2(a)), with an onset speed of $v_{\text{onset}} \approx 1.04 c_s$. The crack state is located in the crack stability phase diagram (Fig.2(c)) at ($G/\Gamma_0$, $\ell_0/\xi$)=(6.65, 0.57).

Movie_3: This animation shows a crack branching scenario (see the third crack pattern in Fig. 2(a), rendered by phase field) in which the pre-strain, i.e., the crack driving force, is increased compared to the previous crack oscillation state. The crack first undergoes oscillations, followed by tip splitting, which is located in the crack stability phase diagram (Fig.2(c)) at ($G/\Gamma_0$, $\ell_0/\xi$)=(7.82, 0.57).

Movie_4: This supershear crack state is located in the crack stability phase diagram (Fig.2(b)) of strain stiffening materials at ($G/\Gamma_0$, $\ell_0/\xi$)=(17.65, 1.33). The crack exhibits a 'wedge-like' shape with a limiting crack velocity of $v_{\text{limit}} \approx 1.44 c_s$. A snapshot of this movie is presented in the third crack pattern in Fig. 4(d).